\documentstyle[11pt,emulateapj]{article}

\def\water{H\( _2 \)O}               % \water H2O
\def\hmol {H\( _2 \)}                % \hmol  H2
\def\xieff {\xi_{\mathrm{eff}}}      % \xieff      xi_eff
\def\xoh {x_{\mathrm{OH}}}           % \xoh        x_OH
\def\hho {\mathrm{H}_2\mathrm{O}}    % \hho        H2O
\def\oh {\mathrm{OH}}                % \oh         OH
\def\hh {\mathrm{H}_2}               % \hh         H2
\def\nh {{n_{\mathrm{H}}}}           % \nh         n_H
\def\ee #1 {\times 10^{#1}}          % \ee p       10^p
\def\ut #1 #2 { \, \mathrm{#1}^{#2}} % \ut unit p  unit^p  
\def\u #1 { \, \mathrm{#1}}          % \u unit     unit
  
\def\kms {\,\mathrm{km\,s}^{-1}}     % \kms        km s^-1
\def\persec {\, \hbox{s}^{-1}}       % \persec     s^-1
\def\percc {\,\mathrm{cm}^{-3}}      % \percc      cm^-3
\def\micron {\, \mu \hbox{m}}
\slugcomment{Submitted to Ap. J. Lett.}

\lefthead{Wardle, Yusef-Zadeh \& Geballe}
\righthead{OH(1720 MHz) masers \& supernova remnants}

\begin{document}

\title{A model for OH(1720 MHz) masers associated with supernova 
remnants, and an application to Sgr A East}

\author{Mark Wardle}
\affil{Special Research Centre for Theoretical Astrophysics, University
of Sydney, NSW 2006, Australia}
\medskip 

\author{Farhad Yusef-Zadeh}
\affil{Department of Physics and Astronomy, Northwestern University, 
Evanston, IL 60208}

\and

\author{T.R. Geballe}
\affil{Joint Astronomy Centre, 660 N. A'ohoku Pl., Hilo, HI 96720}

\begin{abstract} %\section{Abstract}
OH(1720 MHz) masers unaccompanied by 1665/7 MHz line masers have recently 
been proposed as indicators of the interaction of supernova remnants 
(SNRs) and molecular clouds.  We present a model for the masing region 
in which water produced in a C-type shock wave driven into the 
molecular cloud is dissociated as a result of the X-ray flux from the 
SNR.  We note that the magnetic field strengths inferred from 
Zeeman splitting of the 1720 MHz line measure the internal pressure of 
the supernova remnant.

In addition, we discuss the interaction of Sgr A East, a SNR 
candidate, with the 50 km/s cloud at the Galactic Centre and present 
near-infrared observations of \hmol\ emission towards the regions 
where OH(1720 MHz) maser emission is concentrated.  The magnetic field 
strength obtained from earlier Zeeman measurements is 
consistent with rough pressure equilibrium between the postshock gas 
and the X-ray gas filling Sgr A East detected by ASCA. Further, the 
intensity of the v=1--0 S(1) line of \hmol\ is consistent with the 
shock strength expected to be driven into the molecular gas by this 
pressure.  The relative intensities of the \hmol\ lines in Sgr A East 
imply mainly collisional excitation.

\end{abstract}


\section{Introduction}

OH masers have generally been used as a diagnostic for HII regions and 
evolved stars.  However, a recent study by Frail, Goss \& Slysh (1994) 
revealed that the 1720 MHz transition of OH maser emission, when 
unaccompanied by the 1665 and 1667 MHz OH lines, can be an effective 
indicator of shock waves interacting with molecular clouds, 
particularly for supernova remnants (SNRs).  Frail et al.  (1994) 
detected 26 distinct OH(1720 MHz) maser spots along the interface 
between the SNR W28 and an adjacent molecular cloud.  
Evidence of the association between the molecular cloud and W28 comes 
from the distribution of the molecular material following the eastern 
edge of the supernova shell (Wootten 1977).  There are more than a 
dozen Galactic sources and three extragalactic sources in which 1720 
MHz OH masers have been found interior to SNR's with adjacent 
molecular clouds (Yusef-Zadeh, Uchida \& Roberts 1995; Frail et al. 
1996; Yusef-Zadeh et al. 1996; Green et al. 1997; Seaquist, Frayer \& 
Frail 1997).  These masers are close both in position and velocity to 
the interfaces between the remnants and the clouds.  In addition, in 
several of these sources, CO emission lines reach a maximum in both 
brightness and linewidth at the interface between remnant and cloud 
(Wootten 1977).  These are strong observational indications that the 
shocks are caused by the remnants expanding into their respective 
adjacent molecular clouds.

Further support is provided by theoretical studies of the pumping of 
the OH maser lines (Elitzur 1976; Pavlakis \& Kylafis 1996a,b).  The 
1665 and 1667 MHz masers are pumped by far-infrared 
radiation and are therefore associated with HII regions and 
evolved stars.  The OH(1720 MHz) maser is collisionally pumped in 
molecular gas at temperatures and densities between 15-200 K and 
\(10^4-10^6\) cm\(^{-3}\), respectively.  Thus in the absence of the 
1665/7 MHz transitions, the OH 1720 MHz line presumably traces 
cooling, shocked gas. 

However, shock chemistry predicts that OH is \emph{not} abundant in 
the postshock gas as it is rapidly converted to \water\ within the 
shock front.  OH masers adjacent to compact HII regions are produced 
by photodissociation of \water\ (Elitzur \& de Jong 1978;
Hartquist \& Sternberg 1991; Hartquist et al. 1995), but 
in that case there is a strong dissociating FUV flux from the star.  
The dissociating flux is largely absorbed and reradiated in the FIR by 
grains, providing a sufficient IR background to also pump the 1665/7 
MHz transitions (e.g. Pavlakis \& Kylafis 1996b).  This cannot be the 
case for the unaccompanied 1720 MHz masers associated with 
SNR-molecular cloud interactions.

Here we propose that it is the weak X-ray flux from the SNR interior 
that is ultimately responsible for the dissociation of \water\ in the 
shocked molecular gas.  We note that if the OH maser arises in the 
postshock gas, then Zeeman measurements determine the magnetic field 
strength in the postshock gas, and thus measure the pressure within 
the SNR more directly than other methods.  The observations of the IC 
443, W28 and W44 are consistent with this scenario.  Finally, we apply 
these ideas to the interaction of the Galactic center nonthermal 
source Sgr A East, which is either a SNR or a multiple-SNR driven 
bubble, with a molecular cloud at the Galactic center.

\section{A model for the production of OH}

Although the pumping conditions for the 1720 MHz OH masers suggest 
that the masers are associated with shock waves, this is not an 
immediate consequence of shock chemistry (e.g.  Draine, Roberge \& 
Dalgarno 1983; Hollenbach \& McKee 1989).  In J-type shock waves 
molecules are destroyed within the shock front and reform well 
downstream of the hot, ionised gas immediately behind the shock.  OH 
exists as an intermediary in the incorporation of oxygen into water, 
at a temperature of roughly 400 K, too high to explain the 1720 MHz 
masers.  C-type shocks are non-dissociative, and efficiently convert 
the atomic and diatomic oxygen in the unshocked gas to \water\ if the 
shock velocity exceeds 10 \( \kms \) (Draine et al. 1983; Kaufman \& 
Neufeld 1996).  Once the gas temperature exceeds about 400 K, OH is 
rapidly formed by the endothermic reaction
\begin{equation}
	\mathrm{O} + \hh \longrightarrow \oh
	\label{eq:oh_formation}
\end{equation}
but is even more rapidly converted to water by the less-endothermic 
reaction
\begin{equation}
	\oh + \hh \longrightarrow \hho.
	\label{eq:water_formation}
\end{equation}
Thus shock waves do not of themselves produce significant columns of 
OH. 

The production of a significant abundance of OH in the postshock gas 
requires dissociation of \water.  Although this can be achieved by 
intense UV irradiation of the molecular gas (Hartquist \& Sternberg 
1991), the resultant grain heating 
generates a FIR continuum capable of pumping the 1665 and 1667 MHz 
transitions (Pavlakis \& Kylafis 1996b) and cannot explain the 
OH(1720 MHz) masers 
associated with supernova remnants.  In addition, the emission from 
the molecular clouds associated with the remnants do not show 
signatures of UV heating (e.g. Burton et al. 1990; Reach \& Rho 1996).

However, dissociation can occur because of the irradiation of the 
molecular cloud by X-rays produced by the hot gas in the interior of 
the adjacent SNR. The dissociation occurs indirectly: electrons are 
photoejected by the X-rays and collisionally excite the Werner Ly-\( 
\alpha \) band of \hmol.  The subsequent radiative decay contributes 
to a secondary dissociating FUV radiation field.  Irradiation by 
X-rays is more efficient at heating, ionising and dissociating the gas 
rather than heating grains as the ratio of the grain to molecule 
absorption cross-sections for X-rays is much lower than for UV photons 
(Maloney, Hollenbach \& Tielens 1996).  X-rays are much more 
penetrating than UV, and the dissociating flux, although weak, is 
generated throughout the cloud.

\begin{figure*}
\centerline{\epsfxsize=15cm \epsfbox{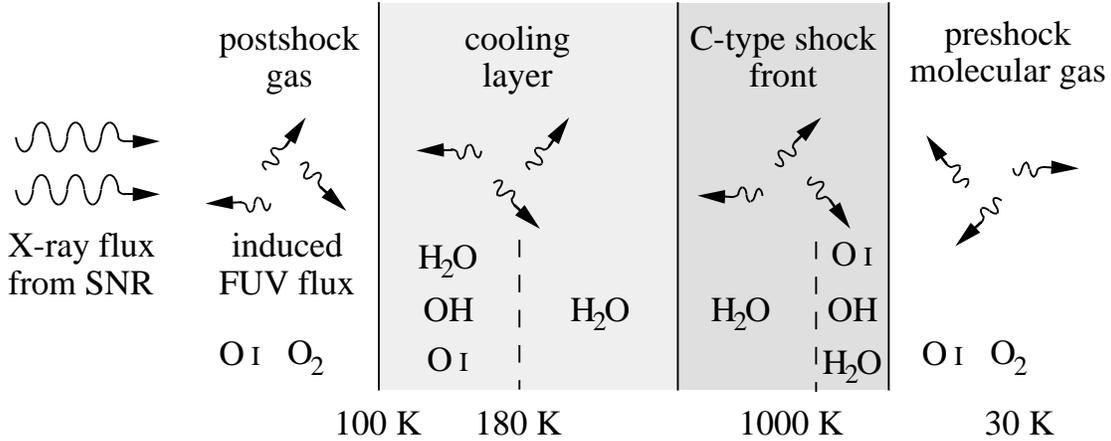}}
\caption{A sketch of the model for the production of OH 
in molecular clouds associated with supernova remnants.  A C-type 
shock is driven into a molecular cloud adjacent to a SNR. The X-ray 
flux from the SNR interior (to the left) permeates the cloud, inducing 
a weak secondary FUV flux that is produced locally throughout the 
cloud.  The shock efficiently wave incorporates atomic and molecular 
oxygen into water.  Once the the shocked gas cools below 180 K it is 
subsequently dissociated to OH and then to OI by the secondary UV flux 
(see text).\label{fig:model}}
\end{figure*}

This suggests a model in which water forms within a C-type shock front 
and is subsequently dissociated by the secondary FUV flux permeating 
the cloud as the shocked gas cools behind the shock front, as sketched 
in Fig.  \ref{fig:model}.  The upper portion of the figure shows the 
velocity field and magnetic field lines for a perpendicular C-type 
shock being driven into a molecular cloud adjacent to the SNR. The 
distinct physical regions delineated by the species containing the 
oxygen that is not locked up in CO are indicated in the lower part of 
the Figure.  The preshock gas, at the right of Fig.  \ref{fig:model} 
is at rest, and the oxygen that is not bound in CO exists in atomic or 
diatomic form.  As the gas is accelerated, compressed, and heated 
within the shock front, the atomic and molecular oxygen is rapidly 
converted to \water\ by reactions (\ref{eq:oh_formation}) and 
(\ref{eq:water_formation}) (Draine et al. 1983; Kaufman \& Neufeld 
1996).  The peak temperature within the shock is typically of order 
1000 K. The shocked gas cools as it drifts behind the shock front.  
The entire structure is subject to the weak dissociating flux produced 
by the X-rays permeating the cloud.  At high temperatures the 
dissociation cannot compete with the OH and \water\ formation 
reactions (\ref{eq:oh_formation}) and (\ref{eq:water_formation}).  
However, the temperature behind the shock eventually drops to the 
point where the reaction rates for formation are so slow that 
dissociation proceeds, and there is a warm layer which is rich in OH.

The analogous scenario for J-type shock waves fails because the water 
abundance \( n(\)\water\( )/\nh \) in the cooling postshock gas exceeds \( 
10^{-5} \) only for preshock densities in excess of \( 10^6\percc \) 
(Elitzur, Hollenbach \& McKee 1989), and the emission from the 
immediate postshock gas is largely absorbed and reradiated in the FIR 
by grains (Hollenbach \& McKee 1989), creating a radiation field that 
is strong enough to pump the OH 1665/7 MHz transitions.

Here we show that the X-ray flux from the SNR incident on the 
water-rich gas behind a C-type shock wave produces a sufficient column 
of OH at the correct temperature and abundance to satisfy the pump 
conditions.  The physical state of X-ray irradiated gas is determined 
by the local heating rate, which can be represented by the parameter
\begin{equation}
	\xieff \approx 1.3\ee -6 \frac{100 F_X}{n_5 N_{22}}
	\label{eq:xi_eff}
\end{equation}
(Maloney et al. 1996) where \( F_X \) is the flux incident on the cloud 
surface (\( \u erg \ut cm -2 \persec \)), \( \nh = n_5 \, 10^5 \percc 
\) is the gas density and \( N_{\mathrm{H}} = N_{22}\,10^{22}\ut cm -2 
\) is the attenuating column to the cloud surface.  \( F_X \approx L_X 
/ 4\pi R^2 \approx 0.01 L_{36}R_{\mathrm{pc}}^{-2}\), where \( L_X = 
L_{36}\, 10^{36} \u erg \persec \) is the SNR X-ray luminosity, and \( 
R_{\mathrm{pc}} \) is the SNR radius in pc, so typically \( F_X \sim 
0.01 \u erg \ut cm -2 \persec \), and \( \xieff \approx 10^{-6} \).  

The dissociation rates for \water\ and OH are
\begin{equation}
	R_{\hho} \approx 8.4 \ee -12  n_5 
	\left(\frac{\xieff}{10^{-6}}\right) \persec
	\label{eq:H20_dissn_rate}
\end{equation}
and
\begin{equation}
	R_{\oh} \approx 4.4 \ee -12  n_5 
	\left(\frac{\xieff}{10^{-6}}\right) \persec
	\label{eq:OH_dissn_rate}
\end{equation}
per molecule respectively (Maloney et al. 1996).  Thus the dissociation 
timescale for \water\ is \( \sim 10^{10} \u s \), comparable to the 
cooling time scale for the gas behind a C-type shock wave.  If 
the temperature is too high the dissociation of \water\ is 
counteracted by rapid reformation via reaction 
(\ref{eq:water_formation}) which has a rate per OH molecule
\begin{equation}
	k(T) = 6.5\ee -13\, n_5\, T^{1.95}\,\exp \left(\frac{-1420}{T}\right) \, \persec \,.
	\label{eq:h20_formation_rate}
\end{equation}
where \( T \) is in Kelvin (Wagner \& Graff 1987).
The dissociation of \water\ does not proceed in the postshock gas until the 
temperature drops to a value \( T_{\mathrm{OH}} \) for which
\( k \approx R_{\hho} \), that is
 \begin{equation}
 	T_{\mathrm{OH}} \approx  175 \left[ 1 - 
 	0.28 \log\left(\xieff / 10^{-6}\right)\right]^{-1} \u K \,.
 	\label{eq:T_OH}
 \end{equation} 
This critical temperature is  weakly-dependent on the X-ray flux because 
of the temperature sensitivity of reaction (\ref{eq:water_formation}).
Once this point is reached, the dissociation of \water\ to 
OH and subsequently to OI proceeds.  As the dissociation rate of OH is 
roughly half that of \water\, there will be a column of OH with \( 
\xoh \approx 10^{-4}\) -- \( 10^{-5} \) and \( T \approx 100 \)--\( 
200\u K \) extending over a distance of about \( 10^{15}\u cm \).  

This model naturally produces the correct conditions for the 
collisional pumping of the OH 1720 MHz transition when SNRs interact 
with molecular clouds.  It implies that Zeeman observations of the 
1720 MHz transition measure the magnetic field in the postshock gas.  
As the magnetic pressure dominates the thermal and ram pressures 
behind the C-type shock, and the postshock gas is in near pressure 
equilibrium with the SNR interior, Zeeman measurements give a good 
estimate of the pressure within the SNR.

IC443, W28 and W44 all show evidence of shocked molecular gas (Wootten 
1977,1981; Dickman et al. 1992) and, in the case of IC443 and W44, a 
contribution to line emission from C-type shock waves (Burton et al. 
1990; Wang \& Scoville 1992; Reach \& Rho 1996).  Claussen et al. 
(1997) used Zeeman splitting in the OH(1720 MHz) line to obtaining 
line-of-sight field strengths \( B_\parallel \) of roughly 0.3 
milliGauss for the masers in W28 and W43, and showing that the field 
strength determinations are consistent with the pressures inferred by 
other means, such as the velocity of H\( \alpha \) filaments.  The 
X-ray luminosities for these three SNRs imply incident X-ray fluxes of 
order \( 0.01 \u erg \ut cm -2 \persec \) (Asaoka \& Aschenbach 1994; 
Rho et al. 1994, 1996).

\section{\hmol\ \( 2 \micron \) emission associated with Sgr A East}

The nonthermal radio source Sgr A East is thought to be a SNR lying 
roughly 30 pc behind the Galactic center (Yusef-Zadeh \& Morris 1987; 
Pedlar et al. 1989).  The recent discovery of OH(1720 MHz) masers, 
unaccompanied by maser emission at 1665 and 1667 MHz, at the interface 
of the 50 \( \kms \) molecular cloud (M-0.02-0.07; Mezger at al 1989) 
and Sgr A East provides strong evidence that these two are physically 
interacting with each other (Yusef-Zadeh et al. 1996).  The maser spots 
surrounding the Sgr A East shell have velocity close to the systemic 
velocity of the SNR near 50 km s\(^{-1}\), and Yusef-Zadeh et al. 
argued that the expansion of a supernova into the Sgr A East molecular 
cloud (i.e.  the 50 km s\(^{-1}\) cloud) and cloud-cloud collisions 
are responsible for the OH(1720 MHz) maser emission.

Recently we used the long slit 1-5\(\mu\)m spectrometer, CGS4, at 
UKIRT to search for shocked \hmol\ emission associated with the OH 
(1720 MHz) masers seen towards Sgr A East.  The 2.5\( '' \) wide slit 
of the spectrometer was placed along two prominent regions of Sgr A 
East where the masers are concentrated (see positions A to G in Figure 
1 of Yusef-Zadeh et al. 1996).  Flux calibration was obtained from an 
observation of BS~6310 and wavelength calibration was obtained from OH 
sky lines.  The 1-0 S(1) line (2.122~$\mu$m) was detected toward both 
regions, with a flux in each row of the array of roughly \( 5\ee -19 
\u W \ut m -2 \).  Figure 2 shows spectra that were seen along 
positions A through G. The second row from the top coincides with 
position A OH maser at \(\alpha(1950) = 17^{h} 42^{m} 33.6^{s}, 
\delta(1950) = -29^{0} 00' 09.6''\) at the interface of Sgr A East.  
The top row is 1.2~arcsec SW, and the 3rd - 6th spectra are 1.2, 2.5, 
3.7, and 4.9 arcsec to the NE of the above coordinates.  In the 
individual spectra the 1-0 S(1) lines can be seen, but other lines of 
H$_{2}$ are too weak for them to be securely identified.  The spectrum 
at the bottom of the figure is the average of the six spectra.  In the 
averaged spectrum the 1-0 S(0) (2.223~$\mu$m) and 2-1 S(1) 
(2.248~$\mu$m) lines of H$_{2}$ are evident, as well as H I 
Br~$\gamma$ (2.166~$\mu$m).  The 2-1 S(1) / 1-0 S(1) and 1-0 S(0) / 
1-0 S(1) line ratios are about 1/7 and 1/3 in the average spectrum. 

The intensity ratios of these lines at the Galactic Center are often 
consistent with collisional excitation rather than fluorescence 
(Gatley et al.  1984; Burton \& Allen 1992; Pak et al.  1996), but the 
high density of the molecular gas at the Galactic center and the 
intense UV radiation field in the region allow the line ratios from 
UV-irradiated gas to resemble those of shock-heated gas (Sternberg \& 
Dalgarno 1989).  While the UV radiation field is known to be strong in 
Sgr A West (10\(^6\) G\(_0\)), the UV radiation field associated with 
Sgr A East is likely to be much less than in the inner pc of the 
Galaxy as Sgr A East probably lies roughly 30 pc behind Sgr A West 
(Pedlar et al.  1989) and the maser spots are distributed in an area 
free of the HII regions that would be associated with massive star 
formation.  Given the weak detections of the 2-1 S(1) and 1-0 S(0) 
lines, the ratios of their intensities with that of the 1-0 S(1) line 
are consistent with those produced by collisional excitation (i.e.  
1/10 and 1/4 respectively), although a fluorescent component arising 
in a lower density environment and producing line ratios of \( \sim 
1/2 \) and \( \sim 2/3 \), respectively, may also contribute.

\begin{figure*}
\centerline{\epsfxsize=15cm \epsfbox{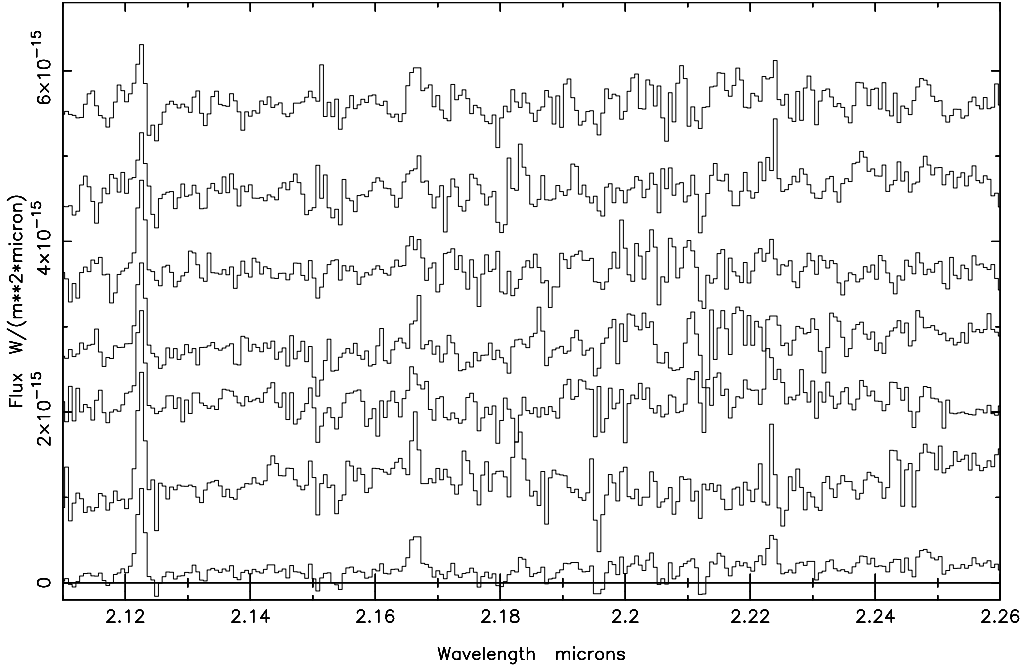}}
\caption{Six adjacent spectra, with vertical scales 
offset from one another, extracted from a CGS4 observation of the 1720 
MHz OH maser spots in Sgr A East and (at the bottom) the averaged 
spectrum of the six.  Each spectrum is of an area $ 1.2'' \times 
2.5'' $.  See text for positions.\label{fig:h2}}
\end{figure*}

\section{Discussion}

We have proposed a model for the OH 1720 MHz masers unaccompanied by 
main-line transitions that are associated with SNRs interacting with 
molecular clouds (Frail et al. 1994, 1996; Green et al. 1997).  A C-type 
shock wave driven into the adjacent cloud produces water that is 
subsequently dissociated by the secondary FUV flux produced by the 
interaction of X-rays from the SNR incident on the molecular cloud.  
The dissociation only becomes significant once the shocked molecular 
gas has cooled to about \( 180 \u K \), naturally producing OH at the 
temperature and abundance required for collisional pumping of the 1720 
MHz transition.  
 
Some consistency checks suggest that this model also applies to the OH 
1720 MHz masers seen towards Sgr A East.  Firstly, within the 
framework of the model the magnetic field strengths 
of roughly 3 mG inferred from Zeeman splitting of the OH masers 
(Yusef-Zadeh et al. 1996) are measurements of the magnetic pressure in 
the postshock gas, which dominates the gas pressure behind the shock 
front.  Equating the magnetic pressure, \( 4\ee -7 \u erg \percc \), 
to \( \rho v_s^2 \), where \( \rho \) is the preshock density and \( 
v_s \) is the shock speed, and adopting a preshock density of \( \nh=2 
\ee 4 \percc \) (Mezger et al.  1989), we infer \( v_s \approx 
25\)--\(30 \kms \).  A C-type shock at this speed produces an 
intensity in the 1-0 S(1) line of \( 10^{-4} \)--\(10^{-3} \u erg \ut 
s -1 \ut cm -2 \) (Draine et al. 1983; Kaufman \& Neufeld 1996).  The 
\emph{measured} intensity, after correcting for extinction assuming that \( 
A_K \approx 2.5 \), is \( 1.5\ee -4 \u erg \ut s -1 \ut cm -2 \).  
Second, the inferred postshock pressure is comparable to the 
pressure of the X-ray emitting gas filling Sgr A East detected 
with ASCA (Koyama et al. 1996), which has \( n_e \approx 6 \percc \) 
and \( T \approx 10 \u keV \), and a pressure of \( 2\ee -7 \u erg 
\percc \).  Finally, we note that the X-ray luminosity from Sgr A 
East is \( 10^{36} \u erg \persec \) (Koyama et al. 1996), providing 
the required dissociating flux for the production of OH behind the 
shock front.

Rho (1995) suggests that centrally-peaked thermal X-ray emission from 
SNRs is characteristic of SNR-molecular cloud interactions.  Green et 
al. (1997) note that all of the SNRs detected in the OH (1720 MHz) line 
that have been observed in X-rays are members of this class.  Our 
model appears to strengthen this link by relying on the X-ray flux 
from the SNR to produce OH, but in principle a cosmic-ray flux 
enhanced by a factor of 100 over the solar neighbourhood value has a 
similar effect on the heating and chemistry of molecular gas (see, 
e.g.  Maloney et al. 1996).  The high-energy cosmic-ray flux in the 
W44, W28 and IC 443 SNRs inferred from EGRET observations is roughly 
two orders of magnitude larger than the solar neighbourhood value 
(Esposito et al. 1996), so a naive extrapolation down to the cosmic-ray 
energies important for ionisation suggests that the cosmic-ray and 
X-ray contributions to the secondary FUV flux may be similar.

In any case, this picture supports the suggestion by Frail et al. 
(1994) that the presence of the OH 1720 MHz line, and the absence of 
the 1665/1667 MHz lines provide a clear diagnostic of shocked 
molecular gas.

\acknowledgements
The CGS4 spectra were obtained as part of the UKIRT Service Programme.  
UKIRT is operated by the Joint Astronomy Centre on behalf of the U.K. 
Particle Physics and Astronomy Research Council.  The Special Research 
Centre for Theoretical Astrophysics is funded by the Australian 
Research Council under the Special Research Centres programme.  F. 
Yusef-Zadeh's work was supported in part by NASA grant NAGW-2518.

\end{document}